\documentclass[a4paper]{article}

\usepackage{INTERSPEECH2022}
\usepackage[acronym]{glossaries}
\usepackage{amsmath, graphicx}
\usepackage{amsfonts}
\usepackage{xcolor}
\usepackage{derivative}
\usepackage{float}

\newacronym{mcd}{MCD}{Mel-ceptral distortion}
\newacronym{mos}{MOS}{mean opinion scores}
\newacronym{ll}{LL}{log likelihood}
\newacronym{rmse}{$\text{RMSE}_{F0}$}{root mean square error of fundamental frequency}
\newacronym{gan}{GANs}{Generative Adversarial Network}
\newacronym{cdf}{CDF}{Cumulative Distribution Functions}

\makeatletter
\newcommand*\bigcdot{\mathpalette\bigcdot@{.5}}
\newcommand*\bigcdot@[2]{\mathbin{\vcenter{\hbox{\scalebox{#2}{$\m@th#1\bullet$}}}}}
\makeatother

\title{FlowVocoder: A small Footprint Neural Vocoder based Normalizing Flow for Speech Synthesis}
\name{Author Name$^1$, Co-author Name$^2$}
\name{Manh Luong$^1$, Viet Anh Tran$^2$}
\address{$^1$Vinai Research\\
  $^2$Deezer Research and Development}
\email{v.manhlt3@vinai.io, vatran@deezer.com}

\begin{document}
%
\maketitle
\begin{abstract}
    Recently, autoregressive neural vocoders have provided remarkable performance in generating high-fidelity speech and have been able to produce synthetic speech in real-time. However, autoregressive neural vocoders such as WaveFlow are capable of modeling waveform signals from mel-spectrogram, its number of parameters is significant to deploy on edge devices. Though NanoFlow, which has a small number of parameters, is a state-of-the-art autoregressive neural vocoder, the performance of NanoFlow is marginally lower than WaveFlow. Therefore, we propose a new type of autoregressive neural vocoder called FlowVocoder, which has a small memory footprint and is capable of generating high-fidelity audio in real-time. Our proposed model improves the density estimation of flow blocks by utilizing a mixture of~\acrfull{cdf} for bipartite transformation. Hence, the proposed model is capable of modeling waveform signals, while its memory footprint is much smaller than WaveFlow. As shown in experiments, FlowVocoder achieves competitive results with baseline methods in terms of both subjective and objective evaluation, also, it is more suitable for real-time text-to-speech applications.
\end{abstract}


\section{Introduction}
Speech synthesis has numerous practical applications in enhancing user experiences such as reading online newspapers and voice assistants. Technically, there are two main stages to produce synthetic speech: generating acoustic features and transforming acoustic features to waveform signals. The second stage usually refers to the vocoder stage. Traditional vocoder approaches generate audio samples using either the spectral density model~\cite{grifinlim} or the basis of the source-filter model~\cite{world}. However, the fidelity of generated speech from those approaches is low and sounds like a robotic voice. Therefore, in recent years, a lot of studies in neural vocoders have provided remarkable performance in producing speech signals from Mel-spectrogram. They are classified into two categories: autoregressive and non-autoregressive vocoders. Regarding high-fidelity, autoregressive vocoders outperform non-autoregressive vocoders. On the contrary, the latter run much faster and are used in real world applications.

For non-autoregressive vocoders, the~\acrfull{gan}-based method is the most successful neural vocoders that provide high-fidelity speech~\cite{melgan,hifigan,parallelWaveGAN,ganTTS,waveGAN,multibandGAN, StyleMelGAN, UnivNet}. Although ~\acrshort{gan}-based vocoders are able to generate synthetic speech in real-time and have a small memory footprint, they are still fragile to train due to adversarial training. To overcome this training issue, MelGAN~\cite{melgan} and HiFi-GAN~\cite{hifigan} utilize an auxiliary loss called features matching between real and generated data to facilitate adversarial training.
Moreover, only relying on adversarial loss causes degradation of audio quality, so the mean square error of Mel-spectrogram~\cite{hifigan} opts to deal with the degradation of synthetic speech.
While Multiband MelGAN~\cite{multibandGAN} replaces the naive features matching with the multi-resolution STFT to better estimate the discrepancy between ground-truth and generated speech. Instead of using auxiliary loss, GAN-TTS~\cite{ganTTS} uses an ensemble of discriminators to take the linguistic features into account and stabilize the training process.

Another vein of research on non-autoregressive vocoders is to leverage normalizing flow models in order to model waveform signals directly from either acoustic features like Mel-spectrogram~\cite{waveglow} or linguistic features~\cite{clarinet}. Both Parallel Wavenet~\cite{parallelwavenet} and Clarinet~\cite{clarinet} are inverse autoregressive flow models, thus, at inference time they are able to synthesize audio in parallel. Those models are trained in the teacher-student procedure from a well trained Wavenet~\cite{wavenet} to benefit from autoregressive models and achieve real-time inference. Nevertheless, they require a well trained Wavenet teacher and a set of auxiliary losses to acquire high fidelity synthesis, thus, they are complicated in training and development. Consequently, flow-based vocoders generally perform worse than autoregressive vocoders with regard to modeling density of audio signal, therefore, their performance is so far behind compared with autoregressive vocoders. 

The second line of research on neural vocoders is autoregressive vocoders~\cite{nanoflow,waveflow,flowwavenet,Yamamoto2019ProbabilityDD} which acquire closet performance compared with ground-truth audio. Despite autoregressive vocoders being capable of modeling the density of audio signals to generate realistic audio, they are immensely slow at inference due to sequential generating. To achieve real-time inference, the authors in~\cite{waveflow} proposed WaveFlow as a general case for both the Wavenet and the WaveGlow model, WaveFlow is able to trade-off between optimizing the likelihood and inference time to synthesize high-quality audio conditioned on Mel-spectrograms. An improved version of Waveflow is Nanoflow~\cite{nanoflow} which operates a sharing density estimator and flow indication embedding to reduce the model's memory footprint.

In this paper, we propose a new autoregressive normalizing flow vocoder called FlowVocoder, which is developed based on sharing a density estimator block~\cite{nanoflow}. Our proposed method has a small memory footprint by sharing a density estimator across flow blocks, and we also enhance the density estimation of flow blocks by using a more flexible transformation function. Also, as described in~\cite{flowpp}, we adopt a new conditioning architecture that is responsible for computing parameters for transformation functions. According to empirical experiments, we modify the conditioning block by removing attention layers since those layers not only do not improve the flexibility of flow blocks but also require more parameters. Consequently, FlowVocoder is able to generate synthetic audio which 
acquired competitive results in terms of both objective and subjective evaluation compared with baseline vocoders.

\section{FlowVocoder}
FlowVocoder is an autoregressive normalizing flow model that enables to generate waveform signals by sampling from a normal Gaussian distribution conditioned on a Mel-spectrogram. This work relies on~\cite{waveflow} that demonstrated an incredible performance of speech synthesis with a very small number of neural network's parameters. In our model, there are two types of mapping: forward mapping and reverse mapping. The reverse mapping $f^{-1}(x)$ is utilized to model the distribution of input data $p(x)$ through a sequence of reverse transformation functions in a simple distribution $p(z)$, while the forward mapping is utilized to generate back an input data $x=f(z)$ from a sampling $z \sim p(z)$. Particularly, in the reserve mapping, the output $z$ is calculated from the input $x$ by going through a sequence of inverse flow layers which correspond to a sequence of coupling transformation functions as $z=f_{k}^{-1}(f_{k-1}^{-1}(...(f_0^{-1}(x))))$, where $f^{-1}_0(x)=x$. Each transformation function is expressed by a mixture of the logarithm of cumulative distribution functions:
\begin{equation}
    f_{k}^{-1}(x_{k-1}) = \sigma^{-1}(\text{MixLogCDF}(x_{k-1}; \pi_k, \mu_k, s_k)).e^{a_k} + b_k
\end{equation}
\begin{equation*}
\begin{split}
    \text{MixLogCDF}(x_{k-1};\pi_k, \mu_k, s_k) = \sum_{i=1}^{M} \pi^i_{k}\sigma((x_{k-1} - \\ \mu^i_{k}).e^{-s^i_{k}})
\end{split}
\end{equation*}
, where $k$ denotes the $k-th$ reverse transformation function. $M$ is the number of logistic mixture components of the cumulative function, while $i$ stands for the $i-th$ component in the logistic mixture function. $\sigma^{-1}$ denotes inverse sigmoid function. The transformation parameters $a_k, b_k, \pi_k, \mu_k, s_k$ of the reverse function $f_k^{-1}(x_{k-1})$ are computed based on the output of the previous function $f_{k-1}^{-1}(x_{k-2})=x_{k-1}=(x_{1,k-1},x_{2,k-1})$, where $x_0$ is input data. We use  a simple block neural network  $g(x_{1,k-1};\theta)$ which outputs five parameters $a_k, b_k, \pi_k, \mu_k, s_k$ for computing transformation parameters, where $g(x_{1,k-1};\theta)$ is a stack of multiple CNN layers and $\theta$ is its parameters. The output $x_{k}=(x_{1,k},x_{2,k})$ is defined as follows:
\begin{equation}
    x_{1,k} = x_{1,k-1}
\end{equation}
\begin{equation}
\begin{split}
    x_{2,k} = & \sigma^{-1}\big (\text{MixLogCDF}(x_{2,k-1};\pi_{k}, \mu_{k}, s_{k}) \big) . e^{a_{k}} + b_{k}
\end{split}
\end{equation}

The above coupling transformation is more expressive than the classic coupling affine transformation, therefore, it is able to increase the flexibility of flow models to capture data density distribution. Moreover, the mixture logistic CDF always has the reversed function due to its monotonic increase property. Also, the Jacobian determinant of this transformation is straightforward to calculate since it is the sum of the probability density function, the derivative of the inverse sigmoid function, and the scale of transformation $a$. The logarithm of Jacobian determinant, denoted as $\log |\text{det}(\mathbf{J}())|$, of the mixture logistic CDF is computed as follow:

\begin{equation}
\begin{aligned}
    \log{|\text{det}(\mathbf{J}(f^{-1}(x)))|} = & \log{|a|} +  \log{\Bigg|\frac{1}{\delta.(1-\delta)} \Bigg|} \\ 
     + &\log{\Bigg| \odv{\sigma^{-1}(\tau)}{\tau} \Bigg|} \\
\end{aligned}
\end{equation}
where
\begin{equation}
\begin{split}
    \delta  & = \text{MixLogPDF}(x; \pi, \mu, s) \\
            & = z - \log(s) - 2.\log(1 + e^{z}) \\
            &\text{where } z = (x-\mu).e^{-s} 
\end{split}
\end{equation}
\begin{equation}
\begin{split}
    \tau & = \text{MixLogCDF}(x;\pi, \mu, s) \\
\end{split}
\end{equation}

Since our model is the autoregressive model, we reshape the 1-dimensional input data $x$ to the 2-dimensional data by splitting $x$ into $H$ groups as $\{X_1, ..., X_H \}$, and then they are stacked to form a 2-D matrix $ X \in \mathbb{R}^{h \times w} $. To transform from the data distribution to the isotropic Gaussian distribution, a sequence of bipartite transformations represented by mixture logistic CDFs are performed recursively to model a conditional dependency between the grouped data
\begin{equation}
    Z_{i} = f^{-1}(X_{<i}, X_{i}; a, b, \pi, \mu, s), \text{where } i= 1,..., H
\end{equation}
. $X_{<i}$ denotes for ${X_1, ..., X_{i-1}}$. The inverse image of $X \in \mathbb{R}^{h \times w}$ on isotropic Gaussian distribution, $Z \in \mathbb{R}^{h \times w}$, is achieved by sequentially performing inverse mapping $f^{-1}: X \mapsto Z$ over rows of input data. For sampling, a sample $Z \in \mathbb{R}^{h \times w}$ is firstly sampled from the isotropic Gaussian, and then it is transformed to the data distribution by autoregressively performing the forward mapping function $f: Z \mapsto X$ over rows of the sampled noise
\begin{equation}
    X_{i} = f(X_{<i}, Z_i; a, b, \pi, \mu, s), \text{where } i=1,...,H
\end{equation}

For each iteration of training, we directly maximize log-likelihood of the data which poses no difficulty to compute by applying the change of variables:
\begin{equation}
\begin{split}
    \log P(X) = & \sum_{i} \bigg( \log |\text{det}(\mathbf{J}(f^{-1}(X_{<i})))| \\ 
    & - \frac{Z_{i}^2}{2} - \frac{\log{2\pi}}{2} \bigg)
\end{split}
\end{equation}
\subsection{Shared density estimator}
\begin{figure}[h!]
    \centering
    \includegraphics[width=.18\textwidth]{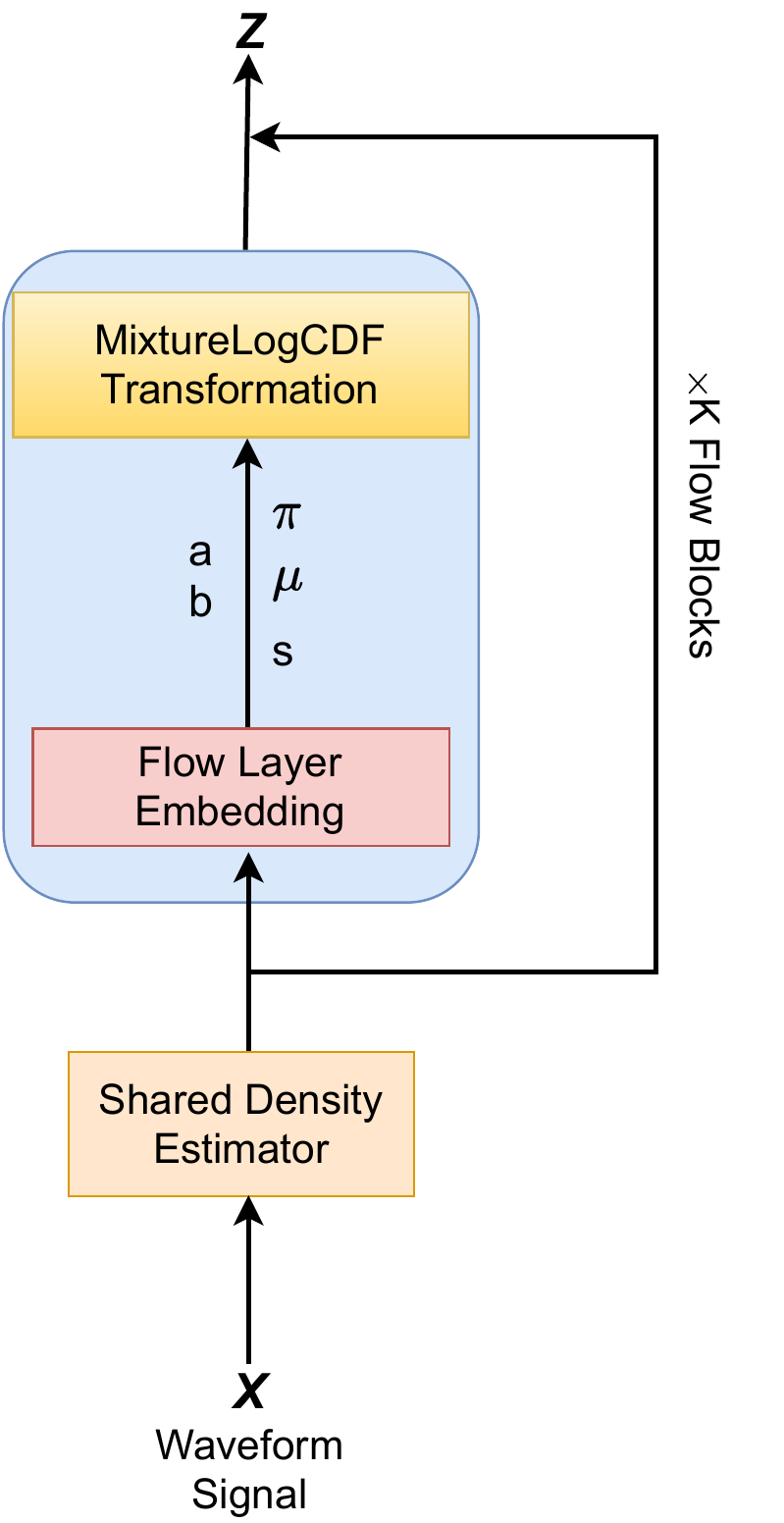}
    \caption{The FlowVocoder architecture with shared density estimator block}
    \label{fig:1}
\end{figure}
In~\cite{nanoflow}, the authors proposed a new model called NanoFlow to reduce the computational and parameter footprint of flow models. As shown in Figure.~\ref{fig:1}, we utilize the same strategy to reduce computation and model size based on sharing a neural density estimator across $K$ flow blocks. Also, a flow indication embedding is used to enable the shared density estimator to be capable of learning multiple contexts for a specific flow block. Consequently, a bipartite transformation block is redefined as:
\begin{equation}
    Z_{i} = f^{-1}(g(X_{<i}, X_{i};\hat{\theta}, \textbf{e}^k); a,b,\pi,\mu,s )
\end{equation}
, where $g(X_{<i}, X_{i};\hat{\theta})$ is the shared density estimator which is utilized for $K$ flow blocks. Embedding vector $ \textbf{e}^k \in \mathbb{R}^D$ represents the $k-th$ flow block, this embedding is fed into the shared density estimator to output proper transformation's parameters. Intuitively, the flow layer embedding $\textbf{e}^k$ as an additional context is able to guide the shared density estimator learning multiple densities with a minimal number of additional parameters.
\begin{figure}[h]
    \centering
    \includegraphics[width=.43\textwidth]{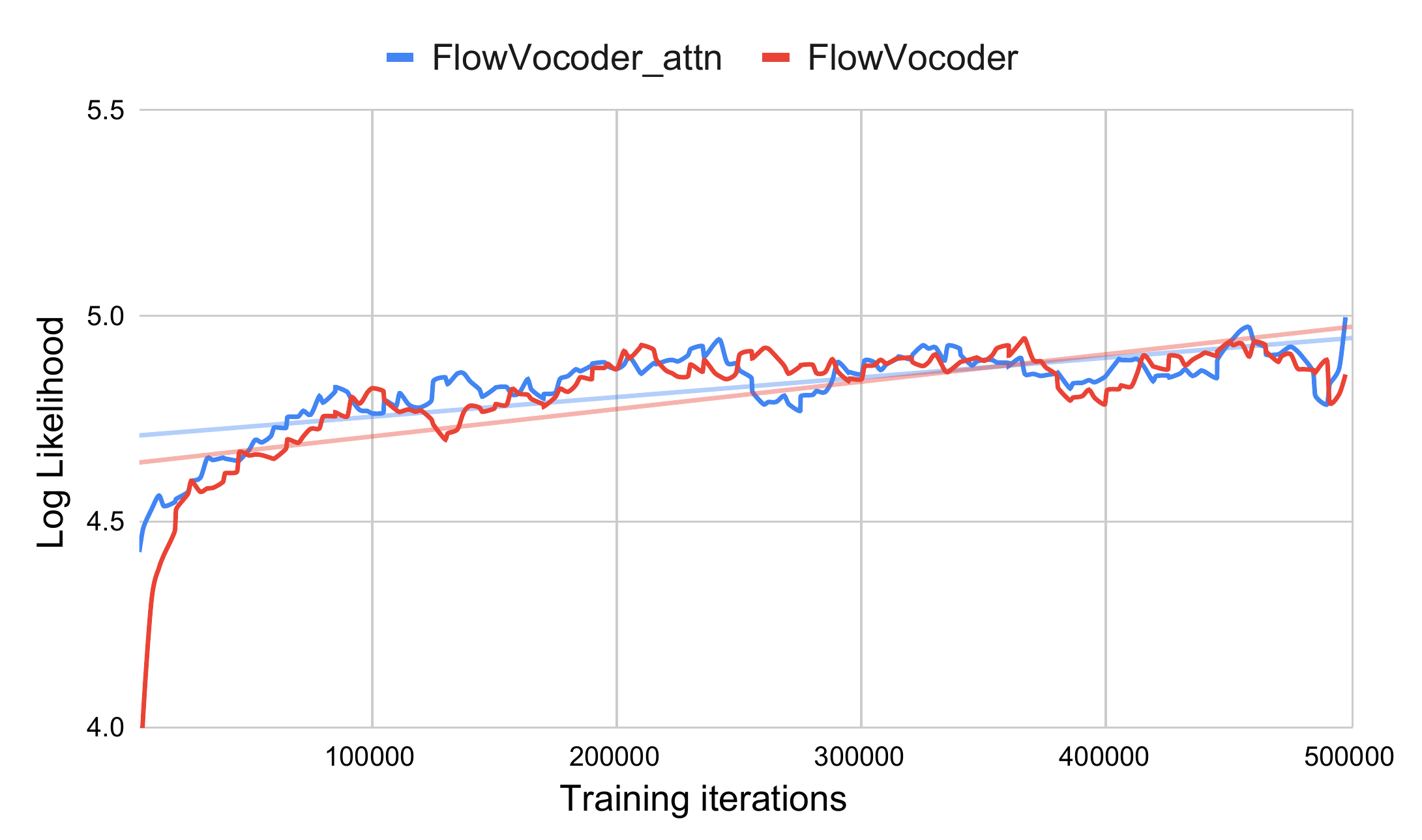}
    \caption{The comparison regarding log likelihood on testing set between FlowVocoder and FlowVocoder with attention layer }
    \label{fig:2}
\end{figure}

As shown in Figure.~\ref{fig:2}, we experimented to compare performance between FlowVocoder using only convolutional layers and FlowVocoder using attention layers as suggested in~\cite{flowpp}. Both models have the same number of residual channels which is 128 channels, and they are trained for 500k iterations. The performances of the two models are estimated on the testing set. It is clear to see that at the beginning of the training process, FlowVocoder with attention surpasses FlowVocoder without attention. However, after enough training, the gap between the two models is insignificant. As a result, flow blocks without attention layers are well-suited for our proposed model. Especially, in the case of deploying on edge devices. Indeed, attention layers require $O(n^2.d)$ computation complexity per layer, therefore, it consumes tremendously computational resources to feed forward. Consequently, the neural block is used to output transformation parameters is: $\text{Conv}_{1\times 1} \xrightarrow{} \text{GatedConv}_{3 \times 3} \xrightarrow{} \text{Conv}_{1 \times 1}$

\subsection{Speech synthesis conditioned on Mel-spectrogram}
After training the FlowVocoder, generating waveform signals conditioned a Mel-spectrogram is straightforward by first sampling a Gaussian noise $z$. Subsequently, since the size of conditioning Mel-spectrograms is mismatched with the sampled noise, we upsample the conditioned Mel-spectrogram to match the size of the noise sample $z$. It is then added as a bias term at each flow block. For mapping a Gaussian noise $z$ forward to waveform signals, the sampled noise is applied to the forward mapping 
\begin{equation}
    X_{i} = \text{MixLogInvCDF}\big (\sigma((Z_{< i} - b).e^{-a}) \big)
\end{equation}
, where $\text{MixLogInvCDF}$ is approximated iteratively by using numerical method. 

\begin{table*}[t!]
\footnotesize
\centering
\caption{The objective evaluation is compared among FlowVocoder and baseline models. The objective metrics are used including the number of model's parameters in million, the mel-ceptral distortion(MCD), the root mean square error of fundamental frequency ($\text{RMSE}_{F0}$), and the log likelihood (LL).}
\label{tab:2}
\begin{tabular}{ccccccc}
\hline
Method   &  Res channels & Parameters(M)$\downarrow$ & MCD$\downarrow$ & $\text{RMSE}_{F0}\downarrow$ &  LL$\uparrow$  \\ \hline
WaveFlow (H=$16$) & $128$  & $22.25$  &  $5.61\pm0.031$ & $30.33\pm0.78$  &  $5.001$ \\
NanoFlow (H=$16$) & $128$  & $2.85$  &  $5.54\pm0.047$ & $38.14\pm0.65$  &  $4.970$ \\
Proposed (H=$16$)& $128$  & $4.14$  &  \textbf{5.37}$\pm$\textbf{0.043} & \textbf{28.25} $\pm$\textbf{0.88}  &  \textbf{5.011} \\  
\hline
\end{tabular}
\end{table*}
\section{Experiment}
We trained and evaluated our proposed model and baseline models on the LJSpeech dataset~\cite{ljspeech}. This dataset contains 13,100 audio clips from a single female speaker, the speech data is approximately 24 hours with a sampling rate of 22.05 kHz. The dataset is split into two parts: 90\% of data for training and 10\% of data for testing. For each utterance, we randomly draw a chunk of 16,000 samples, the chunk of audio is then normalized in the range from 0 to 1 by dividing the maximum waveform value $\text{v}=32768.0$. Next, 80 band log-scale Mel-spectrograms are extracted from the normalized chunk using FFT of size 1024, hop length of size 256, window of size 1024, and Hamming window. To use Mel-spectrograms as a conditioner for speech synthesis, we upsample Mel-spectrogram 256 times by operating two 2D-transpose CNN layers with filter size of $k=[32, 3]$.

Our model and baseline models are trained on 2 Nvidia V100 GPUs with a batch size of 2 for 1M iterations. We use Adam optimizer with initial learning rate of $2 \times 10^{-4}$, and we anneal the learning rate by half for every 200k iterations. The dimension of flow layer embedding is $D=512$ for $\textbf{e}^k \in \mathbb{R}^D$ in the eight-flows block model, and we use $H=16$ to squeeze input audio into 2-D matrix $X \in \mathbb{R}^{h \times w}$.
Our source code and audio samples can be found in our github\footnote{https://v-manhlt3.github.io/FlowVocoder-demo-pages/}

\subsection{Objective evaluation}
\begin{figure}[h!]
    \centering
    \includegraphics[width=0.5\textwidth]{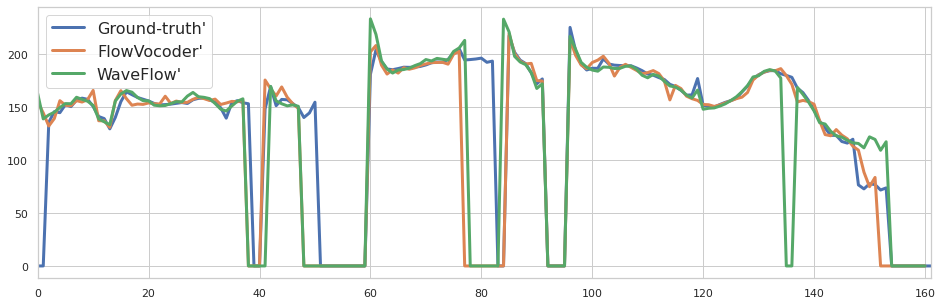}
    \caption{F0 contour of utterance "in being comparatively modern" of FlowVocoder and WaveFlow}
    \label{fig:3}
\end{figure}

For objective evaluation, we draw randomly 100 utterances from the testing dataset to calculate objective metrics. Next, we acquire the performance of synthetic speech with regard to three metrics:~\acrfull{mcd}~\cite{mcd},~\acrfull{ll}, and~\acrfull{rmse}~\cite{rmsef0}. Equations for ~\acrshort{mcd} and ~\acrshort{rmse} calculation are defined as follows:
\begin{equation}
    \text{MCD[dB]} = \frac{10}{\log{10}}\sqrt{2\sum_{m=1}^M (c_r(m) - c_s(m))^2}
\end{equation}
\begin{equation}
    \text{\acrshort{rmse}}\text{[cent]} = 1200\sqrt{(\log_2{(F_r)} - \log_2{(F_s)})^2}
\end{equation}
, where $\text{c}_r$ and $\text{c}_s$ denote Mel-spectrum of groundtruth and synthetic waveform signals, respectively. $M$ denotes the order of Mel-spectrum, $\text{F}_r$ and $\text{F}_s$ denote aperiodic components of groundtruth and synthetic signal.

Table.~\ref{tab:2} shows comparisons between baseline systems and the proposed system regarding objective metrics. NanoFlow is the smallest one regarding the size of models, and the size of our proposed is bigger than NanoFlow by one and a half of the number of parameters. However, the proposed model outperforms Nanoflow in terms of ~\acrshort{mcd} and log likelihood. Especially, FlowVocoder achieves the least Mel-ceptral distance with ground-truth audio, thereby, FlowVocoder is able to produce synthetic speech which has the highest audio quality compared to ones generated by other baselines. In comparision with WaveFlow, our model is much smaller than WaveFlow, about five times smaller in terms of model's size; furthermore, FlowVocoder surpasses WaveFlow with regard to~\acrshort{mcd}, log likelihood, and~\acrshort{rmse}. As shown in Figure.~\ref{fig:3}, FlowVocoder matches $F0$ contour of ground-truth better than WaveFlow, therefore, the generated speech from FlowVocoder is more natural than the generated speech from WaveFlow. This result is also reflected in the subjective evaluation. Finally, we computed the real-time factor(RTF) at the inference step in a single V100 GPU shown in Table.~\ref{tab:rtf}. Although FlowVocoder takes the longest time to synthesize a second of speech, it is still able to generate speech in real-time. The explanation for that phenomenon is due to the burden of approximation of $\text{MixLogInvCDF}$ function in Eq. (11).

\begin{table}[h]
\centering
\caption{The real-time factor(RTF) on a single GPU at the inference step among FlowVocoder and baseline models.}
\label{tab:rtf}
\begin{tabular}{c c c}
\hline
Method   & RTF$\downarrow$ \\ \hline
WaveFlow    &   0.165     \\
NanoFlow    &   0.130     \\
Proposed    &   0.514    \\
\hline
\end{tabular}
\end{table}

\subsection{Subjective evaluation}
For subjective evaluation, we acquired an evaluation of the proposed model and baseline models in terms of naturalness using 5 scales ~\acrfull{mos}, in which 15 participants are asked to assess the quality of synthetic audio. Each participant wears the same headphone when doing the experiment. They are asked to listen to 20 audio clips from each system, then assess for the naturalness of audio on a scale of 1 to 5 with a 1.0 point increments. Since all systems are capable of synthesizing high perceptual audio, we conducted this experiment by asking all participants to listen to the synthetic speech twice. They then give two opinion scores, subsequently, we average these scores to achieve one evaluation sample. Totally, we collected approximately 1,500 evaluation samples, the results of this experiment are represented in the first part of Table.~\ref{tab:1}. All reported ~\acrshort{mos} have statistical significance with $p<0.05$. Our proposed model surpasses both WaveFlow and NanoFlow in terms of high fidelity of audio generating from ground-truth mel-spectrogram. That subjective result points out that the density estimation performance of FlowVocoder is superior with both baselines.  

\begin{table}[h]
\centering
\caption{Subjective evaluation regarding ~\acrshort{mos} with 95\% confidence interval on 20 random utterances from testing dataset. We use the T-test method to examine the statistical significance of all reported MOS with $p<0.05$.}
\label{tab:1}
\begin{tabular}{c c c}
\hline
Method   & MOS$\uparrow$ & 95\%CI \\ \hline
Mel-spectrogram + WaveFlow    &   4.38  & $\pm$ 0.11   \\
Mel-spectrogram + NanoFlow    &   4.18  & $\pm$ 0.12   \\
Mel-spectrogram + Proposed    &   \textbf{4.41}  & $\pm$ \textbf{0.09}   \\
\hline
Tacotron2 + WaveFlow    &   4.26  & $\pm$ 0.11   \\
Tacotron2 + NanoFlow    &   4.07  & $\pm$ 0.13   \\
Tacotron2 + Proposed    &   \textbf{4.30}  & $\pm$ \textbf{0.09}   \\
\hline
ground-truth &   4.62  & $\pm$ 0.08    \\
\hline
\end{tabular}
\end{table}
We also test the quality of synthetic audio combined with Tacotron2~\cite{tacotron2}. The code and pretrained Tacotron 2 model are from the authors github\footnote{https://github.com/NVIDIA/tacotron2}.We test all systems on 20 random utterances from the testing dataset, we first generate Mel-spectrograms using Tacotron 2 and then vocoding these spectrograms by operating baseline vocoders and FlowVocoder. The second part of Table.~\ref{tab:1} reports text to speech experimental results. Regarding FlowVocoder, it acquires competitive performance with NanoFlow and WaveFlow. According to the results, it shows that FlowVocoder is more well-suited for text-to-speech applications.

\section{Conclusions}
In this work, we present a new sort of flow model called FlowVocoder which is capable of modeling audio waveform signals conditioning on Mel-spectrogram. Our proposed model has a small memory footprint, corresponding with fewer model's parameters, by sharing a density estimator across $K$ flow blocks. Moreover, we enhance the flexibility of the bipartite transformation function by using a mixture of CDF. Finally, We modify the conditioning block to calculate transformation parameters efficiently for neural vocoders, particularly, we remove attention layers to be more suitable for deploying on edge devices. As shown in the experiments, FlowVocoder outperforms both WaveFlow and NanoFlow regarding subjective and objective evaluation. Especially, our model matches the $F0$ contour better compared with WaveFlow, therefore, the prosody of generated audio is more natural. Though FlowVocoder acquired impressive results, its inference time should be taken into account. In the future, we could solve the high-latency at inference step by speeding up the approximation of the $\text{MixLogInvCDF}$ function.

\vfill\pagebreak

\bibliographystyle{IEEEtran}

\bibliography{mybib}


\end{document}